\documentstyle[12pt]{article}
\def \beq{\begin{equation}}
\def \eeq{\end{equation}}
\def \ite{{\it et al.}}
\begin{document}
\rightline{EFI 97-33-Rev}
\rightline{Revised January 1998}
\rightline{hep-ph/9707473}
\bigskip

\centerline{\bf IMPROVED TESTS OF RELATIONS}
\centerline{\bf FOR BARYON ISOMULTIPLET SPLITTINGS
\footnote{Revised version submitted to Phys.~Rev.~D.}}
\bigskip
\centerline{\it Jonathan L. Rosner}
\centerline{\it Enrico Fermi Institute and Department of Physics,
University of Chicago}
\centerline{\it 5640 S. Ellis Avenue, Chicago IL 60615}
\bigskip

\centerline{\bf ABSTRACT}
\bigskip

\begin{quote}
The least well-known octet baryon mass is $M_{\Xi^0} = 1314.9 \pm 0.6$ MeV. The
prospect of an improved measurement of its mass by the KTeV experimental
program at Fermilab, and opportunities for improvements in charged and excited
hyperon and $\Delta$ mass measurements, makes it timely to re-examine
descriptions of isospin splittings in baryons containing light quarks. By
examining such relations as the Coleman-Glashow relation $M_n - M_p + M_{\Xi^-}
- M_{\Xi^0} = M_{\Sigma^-} - M_{\Sigma^+}$ one can distinguish between those
models making use of one- or two-body effects involving quarks and those
involving genuine three-body effects.  A hierarchy based on an expansion in
$1/N_c$, where $N_c$ is the number of quark colors, is useful in this respect.
The present status of other quark-model mass relations involving $\Lambda -
\Sigma^0$ mixing and the baryon decuplet is also noted, and the degree to which
one can determine parameters such as quark mass differences and individual
electromagnetic contributions to splittings is discussed. 
\end{quote}
\newpage

\centerline{\bf I.  INTRODUCTION}
\bigskip

The electromagnetic mass splittings of the baryons in the flavor octet of SU(3)
can be understood as a result of several effects in the quark model.  (a) The
$u$ and $d$ quarks have different masses, affecting both static and kinetic
energies. (b) The quarks in a baryon have pairwise Coulomb interactions.  (c)
The strong hyperfine splittings (understood after the advent of quantum
chromodynamics (QCD) as being due to the chromomagnetic interactions between
quarks) can differ as a result of different $u$ and $d$ masses. (d)
Electromagnetic hyperfine interactions between quarks are present.  A sample of
the post-QCD literature, from which earlier observations can be traced, is
contained in Ref.~\cite{PostQCD}. 

Remarkably, there exist plausible limits in which all these effects preserve
one linear relation among the masses of the baryon octet of flavor SU(3)
\cite{CG}: 
\beq \label{eqn:CG}
M_n - M_p + M_{\Xi^-} - M_{\Xi^0} = M_{\Sigma_-} - M_{\Sigma^+}
\eeq
despite substantial symmetry-breaking effects in quark masses.  The resistance
of Eq.~(\ref{eqn:CG}) to symmetry violations was pointed out in
Ref.~(\cite{DGG}), and has been noted recently by Jenkins and Lebed in the
context of a $1/N_c$ expansion \cite{JL}, where $N_c$ is the number of colors
in QCD. (See also the later study by Bedaque and Luty \cite{BL}.) In the
present work we discuss the status and future prospects for testing this
relation, and indicate what might be learned from any violation of it.  We
discuss prospects for improved tests of other relations for isospin-violating
effects, including $\Lambda - \Sigma^0$ mixing and baryon decuplet mass
splittings.  We note the inherent limitations in learning individual terms in
isospin-violating mass differences. 

The stimulus for our re-examination of a 37-year-old problem has come from the
prospect for a substantial improvement in the measurement of the $\Xi^0$ mass
by the KTeV Collaboration at Fermilab. The present value \cite{PDG} is
$M_{\Xi^0} = 1314.9 \pm 0.6$ MeV, while the next most poorly measured mass is
$M_{\Xi^-} = 1321.32 \pm 0.13$ MeV.  It is very likely that KTeV could measure
the $\Xi^0$ mass to comparable or better accuracy, perhaps to $\pm 0.1$ MeV
\cite{KTeV}.  At the same time, a new round of experiments with hyperon beams
\cite{SELEX,HCP} is capable of improving information on $\Xi^-$ and hyperon
resonance masses, while experiments at the Continuous Electron Beam Accelerator
Facility (CEBAF) can improve our knowledge of $\Delta$ resonance isospin
splittings.  The need for such improvement has been stressed recently in
Ref.~\cite{Deltas}.

We begin in Section II with a general discussion of quark-model effects on
isospin-violating mass differences, ending up with a derivation of
(\ref{eqn:CG}) and several other relations.  These are the most general which
follow from the absence of three-body effects \cite{JL}. We discuss the present
and potential experimental situation in Section III, and the degree to which it
is possible to estimate the individual contributions to mass splittings in
Section IV.  We remark on sources of possible violation of Eq.~(\ref{eqn:CG})
and the other relations in Section V, comparing our work with the more general
treatment of Ref.~\cite{JL}.  We comment briefly on charmed baryons in
Section VI, and summarize in Section VII. 
\newpage

\centerline{\bf II.  ISOSPIN VIOLATIONS IN THE QUARK MODEL}
\bigskip

\leftline{\bf A.  Quark mass differences}
\bigskip

The $u$ and $d$ quarks have intrinsic masses which differ by a couple of MeV.
Typical values at scales of 1 GeV \cite{Leut} are $m_u \simeq 5$ MeV$/c^2$,
$m_d \simeq 9$ MeV$/c^2$.  Corresponding estimates for the strange quark mass
range from about 100 to 200 MeV/$c^2$. When quarks are incorporated into
hadrons, more appropriate ``constituent'' values (see, e.g., Refs.~\cite{DGG,
GR,TASI}) are $m_u,m_d = {\cal O}(350)$ MeV/$c^2$, $m_s = {\cal O}(500)$
MeV/$c^2$, with $m_d - m_u$ of order a few MeV$/c^2$ but quite uncertain.  We
shall denote the constituent-quark isospin-violating mass difference by $\Delta
\equiv m_u - m_d$. It will be a free parameter in our description of
isospin-violating baryon mass splittings. 

The quarks' kinetic energies $T$ may also depend on their masses. Without
detailed knowledge of dynamics, it is difficult to anticipate this dependence.
For an effective potential $V = \lambda r^\nu$, the virial theorem $\langle T
\rangle = \langle (r/2)dV/dr \rangle$ implies $\langle T \rangle = \lambda \nu
\langle r^\nu \rangle /2$, while the scaling of the Schr\"odinger equation
\cite{QR} implies $\langle r^\nu \rangle \sim m_Q^{-\nu/(\nu +2)}$.  Thus for a
potential with $\nu < 0$ kinetic energies increase with increasing quark mass,
while for a potential with $\nu > 0$ kinetic energies decrease with increasing
quark mass.  We shall consequently parametrize kinetic energies simply with
labels $K_q$ for those contributions which act as one-body operators and
$K_{q_i q_j}$ for those contributions which depend on interactions with each
individual other quark.

\bigskip

\leftline{\bf B.  Pairwise Coulomb interactions}
\bigskip

Each quark pair in a hadron has a Coulomb interaction energy
\beq \label{eqn:coul}
\Delta E_{ij~\rm em} = \alpha Q_i Q_j \langle \frac{1}{r_{ij}}
\rangle~~~,
\eeq
where $\alpha \simeq 1/137$ is the electromagnetic fine structure constant,
$Q_i$ is the charge of quark $i$ in units of the proton charge, and $\langle
1/r_{ij} \rangle$ is the expectation value of the inverse distance between the
members of the pair. 

In the flavor-SU(3) limit one expects $\langle 1/r_{ij} \rangle$ to be
universal throughout a multiplet.  Thus, for example, every quark pair in every
octet baryon should have the same value of this quantity.  In this limit, we
parametrize the interaction energy $\Delta E_{ij~\rm em} = a Q_i Q_j$, where
$a$ is some universal constant.  We shall explore the possible violations of
this assumption in Sec.~V. 
\bigskip

\leftline{\bf C.  Strong hyperfine interactions}
\bigskip

Quarks are bound in hadrons by a dominantly spin-independent force which
becomes strong at large distances.  In addition, they experience a
spin-dependent force due to gluon exchange which acts dominantly on pairs in an
S-wave state.  For pairs of quarks in a baryon, one has a (strong) hyperfine
interaction energy 
\beq \label{eqn:HFs}
\Delta E_{ij~\rm HFs} = {\rm const.} \frac{|\Psi_{ij}(0)|^2
\langle \sigma_i \cdot \sigma_j \rangle}{m_i m_j}~~~,
\eeq
where 
$|\Psi_{ij}(0)|^2$ is the square of the S-wave wave function of two quarks at
zero relative separation, and the constant is universal for all pairs of quarks
in a baryon. In the limit in which the hyperfine interaction is given by
one-gluon exchange, this constant is of first order in $\alpha_s$.  The nucleon
-- $\Delta$ splitting of about 300 MeV$/c^2$ is an example of a QCD hyperfine
effect. 

We shall assume for the moment that $|\Psi_{ij}(0)|^2$ is universal for all
quark pairs in octet baryons.  We then find a contribution to the hyperfine
energy $\Delta E_{ij~\rm HFs} = b \langle \sigma_i \cdot \sigma_j \rangle/ (m_i
m_j)$. 

The calculation of strong hyperfine splittings in baryons requires evaluation
of $\langle \sigma_i \cdot \sigma_j \rangle$ for each quark pair. Since ${\bf
S} = \sum_i (\sigma_i/2)$, we use the value of ${\bf S}^2$ to evaluate the
sum of $\langle \sigma_i \cdot \sigma_j \rangle$ for all pairs, with the result
\beq \label{eqn:sum}
\langle \sum_{i<j} \sigma_i \cdot \sigma_j \rangle = \left\{
\begin{array}{c} - 3~~~(S = 1/2) \\ + 3~~~(S = 3/2) \end{array}
\right\}~~~.
\eeq

In any color-singlet baryon, Fermi statistics and the antisymmetry of any two
quarks with respect to color interchange lead to symmetry in the remaining
(space $\times$ spin $\times$ flavor) variables.  For ground-state baryons with
two identical quarks (including those involved in the Coleman-Glashow
relation), the two like quarks must hence be in a state symmetric with respect
to spin, i.e., of spin 1, and hence must have $\langle \sigma \cdot \sigma
\rangle = 1$.  For any baryon in the flavor decuplet, such as $\Delta^{++} =
uuu$ (with $S = 3/2$), each pair has this value, consistent with the result
(\ref{eqn:sum}).  For any octet baryon state $q_i q_i q_j~ (j \ne i)$, one then
concludes $\langle \sigma_i \cdot \sigma_j \rangle = -2$. 
\bigskip

\leftline{\bf D.  Electromagnetic hyperfine interactions}
\bigskip

The electromagnetic interaction between quarks in a baryon has a spin-dependent
(hyperfine) contribution 
\beq \label{eqn:HFe}
\Delta E_{ij~\rm HFe} = - \frac{2 \pi \alpha Q_1 Q_2 |\Psi(0)_{ij}|^2
\langle \sigma_i \cdot \sigma_j \rangle}{3 m_i m_j}~~~.
\eeq
Again assuming universality of the wave functions, we can parametrize this
effect as $\Delta E_{ij~\rm HFe} = c Q_i Q_j \langle \sigma_i \cdot \sigma_j
\rangle/(m_i m_j)$. 
\bigskip

\leftline{\bf E.  Summary of effects}
\bigskip

We can now collect all the results for baryon isospin-violating mass shifts
into quantities organized according to the isospin of the splittings.  We
obtain seven $I=1$ combinations, three $I=2$ combinations, and one $I=3$
combination \cite{JL}. 
\bigskip

{\it 1. $\Delta I=1$ splittings.}

$$
N_1 \equiv M_p - M_n = \Delta + K_u - K_d + K_{uu} - K_{dd}
$$
\beq \label{eqn:na}
+ \frac{a}{3} + b \left( \frac{1}{m_u^2} - \frac{1}{m_d^2} \right) +
\frac{c}{9} \left( \frac{4}{m_u^2} - \frac{1}{m_d^2} \right)~~~, 
\eeq
$$
\Sigma_1 \equiv M_{\Sigma^+} - M_{\Sigma^-} = 2 \Delta + 2(K_u - K_d) +
2(K_{us} - K_{ds}) + K_{uu} - K_{dd}
$$
\beq \label{eqn:siga}
- \frac{a}{3} + b \left(\frac{1}{m_u^2} - \frac{1}{m_d^2} + \frac{4}{m_d m_s} -
\frac{4} {m_u m_s} \right) + \frac{c}{9} \left( \frac{4}{m_u^2} -
\frac{1}{m_d^2} + \frac{4}{m_d m_s} + \frac{8}{m_u m_s} \right)~~~, 
\eeq
$$
\Xi_1 \equiv M_{\Xi^0} - M_{\Xi^-} = \Delta + K_u - K_d + 2(K_{us} - K_{ds}) 
$$
\beq \label{eqn:xi}
- \frac{2a}{3} + b \left( \frac{4}{m_d m_s} - \frac{4}{m_u m_s} \right) +
\frac{c}{9} \left( \frac{4}{m_d m_s} + \frac{8}{m_u m_s} \right)~~~, 
\eeq
\beq \label{eqn:slmix}
M_{\Lambda \Sigma^0} = \sqrt{3}\left[ b \left( \frac{1}{m_d m_s}
- \frac{1}{m_u m_s} \right) + \frac{c}{9} \left(\frac{1}{m_d m_s}
+ \frac{2}{m_u m_s} \right) \right]~~~,
\eeq
$$
\Delta_1 \equiv 3 M_{\Delta^{++}} + M_{\Delta^+} - M_{\Delta^0}
- 3 M_{\Delta^-} = 10 \left[ \Delta + K_u - K_d + K_{uu} - K_{dd} \right.
$$
\beq \label{eqn:dela}
\left. + \frac{a}{3} + b \left( \frac{1}{m_u^2}
- \frac{1}{m_d^2} \right) + \frac{c}{9} \left( \frac{4}{m_u^2}
- \frac{4}{m_d^2} \right) \right]~~~,
\eeq
$$
\Sigma^*_1 \equiv M_{\Sigma^{*+}} - M_{\Sigma^{*-}} = 2 \Delta + 2(K_u - K_d) +
2(K_{us} - K_{ds}) + K_{uu} - K_{dd}
$$
\beq \label{eqn:sigsa}
-\frac{a}{3} + b \left( \frac{1}{m_u^2} - \frac{1}{m_d^2} + \frac{2}{m_u m_s} -
\frac{2}{m_d m_s} \right) + \frac{c}{9} \left( \frac{4}{m_u^2} -
\frac{1}{m_d^2} - \frac{4} {m_u m_s} - \frac{2}{m_d m_s} \right)~~~, 
\eeq
$$
\Xi^*_1 \equiv M_{\Xi^{*0}} - M_{\Xi^{*-}} = \Delta + K_u - K_d + 2(K_{us} -
K_{ds}) 
$$
\beq \label{eqn:xis}
- \frac{2a}{3} + 2b \left( \frac{1}{m_u m_s} - \frac{1}{m_d m_s} \right)
-\frac{c}{9} \left( \frac{2}{m_d m_s} + \frac{4}{m_u m_s} \right)~~~. 
\eeq

These quantities are related to one another by
\beq \label{ara}
N_1 = \Sigma_1 - \Xi_1 = \Sigma^*_1 - \Xi^*_1 = \Delta_1/10~~~,
\eeq
\beq \label{eqn:arb}
2 \sqrt{3} M_{\Lambda \Sigma^0} = \Sigma_1 - \Sigma_1^*~~~.
\eeq
The Coleman-Glashow relation (\ref{eqn:CG}) is one of these; the remaining ones
require information on the baryon decuplet.  Eq.~(\ref{eqn:arb}) has been
derived in Ref.~\cite{FLNC}.
\bigskip

{\it 2.  $\Delta I = 2$ splittings.}

$$
\Sigma_2 \equiv M_{\Sigma^+} + M_{\Sigma^-} - 2M_{\Sigma^0} = K_{uu} + K_{dd}
- 2 K_{ud}
$$
\beq \label{eqn:sigb}
+ a + b \left( \frac{1}{m_u} - \frac{1}{m_d} \right)^2 + \frac{c}{9} \left(
\frac{2}{m_u} + \frac{1}{m_d} \right)^2~~~, 
\eeq
$$
\Delta_2 \equiv M_{\Delta^{++}} - M_{\Delta^+} - M_{\Delta^0} + M_{\Delta^-} =
2 \left[ K_{uu} + K_{dd} - 2 K_{ud} \right.
$$
\beq \label{eqn:delb}
\left. + a + b \left( \frac{1}{m_u} - \frac{1}{m_d} \right)^2 + \frac{c}{9}
\left( \frac{2}{m_u} + \frac{1}{m_d} \right)^2 \right]~~~, 
\eeq
$$
\Sigma^*_2 \equiv
M_{\Sigma^{*+}} + M_{\Sigma^{*-}} - 2M_{\Sigma^0} = K_{uu} + K_{dd} - 2 K_{ud}
$$
\beq  \label{eqn:sigsb}
+ a + b \left( \frac{1}{m_u} - \frac{1}{m_d} \right)^2 + \frac{c}{9} \left(
\frac{2}{m_u} + \frac{1}{m_d} \right)^2~~~. 
\eeq
These quantities are all proportional to one another:
\beq
\Sigma_2 = \Delta_2/2 = \Sigma^*_2~~~.
\eeq
The $\Delta I = 2$ relation (\ref{eqn:sigb}) will turn out to be useful, when
combined with the others, in determining the individual contributions to the
mass splittings (Sec.~IV). 
\bigskip

{\it 3.  $\Delta I = 3$ splitting.}
One combination of the $\Delta$ masses vanishes:
\beq \label{eqn:delc}
\Delta_3 \equiv M_{\Delta^{++}} - 3 M_{\Delta^+} + 3 M_{\Delta^0}
- M_{\Delta^-} = 0~~~.
\eeq
This will be useful in eliminating the $\Delta^-$ mass from other relations,
since no value is quoted \cite{PDG} for it.  If (\ref{eqn:delc}) is used, one
finds $M_{\Delta^+} - M_{\Delta^0} = (1/3)(M_{\Delta^{++}} - M_{\Delta^-}) =
N_1$ and $M_{\Delta^{++}} - 2 M_{\Delta^+} + M_{\Delta^0} = M_{\Delta^+} - 2
M_{\Delta^0} + M_{\Delta^-} = \Delta_2/2$.  These relations have been
employed in many of the studies in Refs.~\cite{PostQCD}, \cite{JL},
\cite{BL}, and earlier works quoted by them.
\bigskip

{\it 4.  Discussion.}
We did not need to expand in powers of $m_s - m_d$ or $m_d - m_u$ to obtain the
above relations.  On the other hand, we did assume universality of quark-model
wave functions, i.e., universal values of $\langle 1/r_{ij} \rangle$ and
$|\Psi(0)_{ij}|^2$.  Since the quark masses are arbitrary, the electromagnetic
hyperfine terms automatically will have the same structure as the strong ones,
aside from a weighting of inverse quark masses by quark charges.  As we shall
see in Sec.~V, one can in fact relax the universality assumption, replacing it
by universality of interaction of {\it any given pair} regardless of the baryon
in which it is found. The two-body kinetic terms in fact exhibit this feature.

The relations for the mass splittings are equivalent, upon identification of
terms, to ones which have been obtained previously within the context of
specific models \cite{models,Isgur}. However, as we shall see in Sec.~IV, in
the present approach one is prevented from identifying the magnitude of
individual terms (such as Coulomb and hyperfine self-energies) without making
additional assumptions about the one- and two-body kinetic terms.  The
relations indeed hold under {\em arbitrary} forms of one- and two-body quark
forces. For spin-independent forces this is illustrated by the completely
general nature of the $K_i$ and $K_{ij}$ terms, but it is true when spin is
included as well.  This was, in fact, noted before the advent of QCD
\cite{RSS,JFpre}. 
\bigskip

\centerline{\bf III.  EXPERIMENTAL SITUATION}
\bigskip

\leftline{\bf A.  Present}
\bigskip

\begin{table}
\caption{Masses of baryon octet and decuplet members, in MeV$/c^2$.}
\begin{center}
\begin{tabular}{|c c|c c|} \hline
Octet  &      & Decuplet & \\
Baryon & Mass & Baryon & Mass \\ \hline
$p$ & $938.27231 \pm 0.00028$ &
  $\Delta^{++}$ & $1231.04 \pm 0.17$ \\
$n$ & $939.56563 \pm 0.00028$ &
  $\Delta^+$ & $1234.9 \pm 1.4$ \\
$\Lambda$ & $1115.684 \pm 0.006$ &
  $\Delta^0$ & $1233.77 \pm 0.19$ \\
$\Sigma^+$ & $1189.37 \pm 0.07$ &
  $\Sigma^{*+}$ & $1382.8 \pm 0.4$ \\
$\Sigma^0$ & $1192.55 \pm 0.08$ &
  $\Sigma^{*0}$ & $1383.7 \pm 1.0$ \\
$\Sigma^-$ & $1197.436 \pm 0.033$ &
  $\Sigma^{*-}$ & $1387.2 \pm 0.5$ \\
$\Xi^0$ & $1314.9 \pm 0.6$ &
  $\Xi^{*0}$ & $1531.80 \pm 0.32$ \\
$\Xi^-$ & $1321.32 \pm 0.13$ &
  $\Xi^{*-}$ & $1535.0 \pm 0.6$ \\ \hline
\end{tabular}
\end{center}
\end{table}

The individual masses of members of the baryon octet and decuplet are
summarized in Table 1 \cite{PDG}. 
The measured values of the octet mass splittings are
$$
N_1 = -1.293~{\rm MeV}/c^2~~,~~~
\Sigma_1 = - 8.07 \pm 0.08 ~{\rm MeV}/c^2~~,~~~
$$
\beq \label{eqn:ospl}
\Xi_1 = - 6.4 \pm 0.6 ~{\rm MeV}/c^2~~,~~~
\Sigma_2 = 1.71 \pm  0.18 ~{\rm MeV}/c^2~~~.
\eeq
The left-hand side of the relation (\ref{eqn:CG}) is $7.7 \pm 0.6$ MeV$/c^2$,
consistent with the right-hand side of $8.07 \pm 0.08$ MeV$/c^2$. 

The measured values of the decuplet mass splittings are
\beq \label{eqn:dspl}
\Sigma_1^* = -4.4 \pm 0.64  ~{\rm MeV}/c^2~~,~~~
\Xi_1^* = -3.2 \pm 0.6 ~{\rm MeV}/c^2~~,~~~
\Sigma_2^* = 2.6 \pm 2.1 ~{\rm MeV}/c^2~~~.
\eeq
The relation
\beq \label{eqn:sa}
\Sigma_1 - \Xi_1~(= - 1.67 \pm 0.6 ~{\rm MeV}/c^2) =
\Sigma^*_1 - \Xi^*_1~(= -1.2 \pm 0.9 ~{\rm MeV}/c^2)
\eeq
is satisfied, albeit with large uncertainty. So is the relation
\beq \label{eqn:sb}
\Sigma_2~(= 1.71 \pm  0.18 ~{\rm MeV}/c^2) =
\Sigma^*_2~(= 2.6 \pm 2.1 ~{\rm MeV}/c^2)~~~.
\eeq

In order to compare relations involving $\Delta$ masses, we must use the
vanishing of the $\Delta I = 3$ combination $\Delta_3$ to eliminate
$M(\Delta^-)$.  We then obtain one relation: 
\beq \label{eqn:dsb}
M(\Delta^{++}) - 2 M(\Delta^+) + M(\Delta^0) = \Sigma^*_2~(
= \Sigma_2)~~~.
\eeq
The left-hand side of this expression is $-5.0 \pm 2.8$ MeV/$c^2$, not
particularly consistent with (\ref{eqn:sb}).  We shall see in Sec.~V that the
hierarchy \cite{JL} of mass relations based on the $1/N_c$ expansion moderates
this difficulty by finding Eq.~(\ref{eqn:sb}) to be no more accurate than
$\Sigma_2 = 0$ or $\Sigma_2^* = 0$. On the other hand, Eq.~(\ref{eqn:dsb}) with
$\Sigma^*_2$ on the right-hand side is expected in the $1/N_c$ approach to be
better behaved by an order of magnitude.  It has been noted in
Refs.~\cite{Deltas} that the value quoted for $M(\Delta^+)$ in Ref.~\cite{PDG}
may not be reliable.

If we combine $\Delta_1= 10 N_1$ with $\Delta_3 = 0$, we find $M(\Delta^+) =
N_1 + M(\Delta^0)$. (We shall discuss the accuracy of this relation in Sec.~V.)
We can then substitute in (\ref{eqn:dsb}) to find (see also, e.g., \cite{BL})
\beq \label{eqn:dsa}
M(\Delta^{++}) - M(\Delta^0)~(= - 2.7 \pm 0.3~{\rm MeV}/c^2)
= \Sigma_2^*~(\Sigma_2) + 2 N_1~~~.
\eeq
The right-hand side is $-0.0 \pm 2.1$ MeV/$c^2$ if we use $\Sigma_2^*$
(permitted in Ref.~(\cite{JL}) and $-0.88 \pm 0.18$ MeV/$c^2$ if we use
$\Sigma_2$. 

\bigskip
\leftline{\bf B.  Future}
\bigskip

The KTeV Collaboration at Fermilab \cite{KTeV} has obtained a large sample of
$\Xi^0$'s in a neutral hyperon beam.  The detector is able to observe both
charged and neutral particles in the decay $\Xi^0 \to \pi^0 \Lambda \to \pi^0 p
\pi^-$.  Reasonable prospects exist for reducing the error on $M_{\Xi^0}$ to
$\pm 0.1$ MeV/$c^2$. 

The next most poorly known mass is that of the $\Xi^-$.  Experiments in a new
charged hyperon beam at Fermilab \cite{SELEX,HCP} could reduce the errors on
this quantity. 

The prospects are thus good for reducing the error on the test of the
Coleman-Glashow relation (\ref{eqn:CG}) by at least a factor of 6, to 0.1
MeV$/c^2$ or less.  This is comparable to the accuracy to which the relation is
expected to hold, according to the analysis of Ref.~\cite{JL}. 

The relation (\ref{eqn:arb}) predicts $M(\Lambda \Sigma^0) = - 1.06 \pm 0.19$
MeV/$c^2$.  A test requires one to measure the isospin impurity of the
$\Lambda$ (or, more difficult, of the $\Sigma^0$).  One conceivable way to do
this would be to study the deviations from apparent charge-independence in the
decays $\Sigma^* \to \pi \Lambda$, taking careful account of phase-space
differences and electromagnetic final-state interactions.  One would need to
measure the widths of $\Sigma^{* \pm}$ to a percent, beyond present accuracy. 

The other relations derived above require reduction of errors on the decuplet
masses.  Perhaps the best prospects in this respect involve the combinations
$\Sigma^*_{1,2}$ and $\Xi^*_1$, for which improved values could be obtained in
charged hyperon beams at Fermilab \cite{SELEX,HCP}.  Studies of $\gamma p \to
\Delta^+ \to \pi^0 p$, for example at the Continuous-Beam Electron Facility
(CEBAF), could in principle reduce the error on $M(\Delta^+)$. 
\bigskip

\centerline{\bf IV.  INDIVIDUAL TERMS}
\bigskip

With assumptions about quark masses \cite{GR} and kinetic one- and two-body
terms $K_{q_i q_j}$, one can evaluate individual terms in the expressions for
the mass splittings, such as the quark mass difference $\Delta = m_u - m_d$ and
the Coulomb and electromagnetic hyperfine terms, in a model-independent way. 
[In several earlier studies, dynamical models permitted estimates of the
magnitude of the kinetic one- and two-body terms \cite{models} and hence of
$\Delta$.]  If one does not estimate kinetic terms, the number of parameters is
too large to permit a model-independent evaluation of individual terms. 

Among the seven $\Delta I = 1$ splittings noted in Section II, there are four
relations, so only three are independent.  We may take these as $N_1$,
$\Sigma_1$, and (for example) $\Sigma^*_1$.  Among the three $\Delta I = 2$
splittings there are two relations, so we may take the best-known one
($\Sigma_2$) as independent. If we neglect the kinetic two-body terms, we have
four experimental quantities with which to determine the five quantities
$\Delta$, $K_u-K_d$, $a$, $b$, and $c$, given estimates of the nonstrange quark
mass $\bar m \equiv (m_u + m_d)/2$ and the strange quark $m_s$. 

Each of the three $\Delta I = 1$ splittings contains the same combination
$\Delta + K_u - K_d$.  Thus, if we were not concerned with the individual
values of $\Delta$ and $K_u - K_d$, we could hope to use one of the $\Delta I =
1$ splittings to eliminate $\Delta + K_u - K_d$, leaving three parameters
(essentially $a$, $b \Delta$, and $c$) to determine with the help of three
experimental numbers. 

A fit to overall octet and decuplet masses \cite{GR} leads to $\bar m = 363$
MeV/$c^2$, $m_s = 538$ MeV/$c^2$, and hence $r \equiv \bar m/m_s = 0.675$. 
The same fit also implies $\beta \equiv b/\bar m^2 = 50$ MeV/$c^2$ (as
determined, for example, by the splitting between nucleon and
$\Delta$ states). Thus we might hope to extract a value of $\Delta$
from $b \Delta$ as determined above, and then learn $K_u - K_d$ as well. 

Let us neglect the $u-d$ mass difference in $c$ terms, and note that the $b$
term in $\Sigma_2$ is of second order in $\Delta$.  Then we may write
$\Sigma_2 = a + \gamma$ and
\beq \label{eqn:b}
\Sigma_1 - 2 N_1 + \frac{2}{3} \Sigma_2 = - \frac{a}{3} + 2 \beta \frac{\Delta}
{\bar m}(1 + 2 r) + \gamma (1 + 4r)/3~~~,
\eeq
\beq \label{eqn:c}
\Sigma_1^* - 2 N_1 + \frac{\Sigma_2}{3} = - \frac{2a}{3} + 2 \beta
\frac{\Delta} {\bar m} (1 - r) - 2 \gamma r/3~~~, 
\eeq
where $\gamma \equiv c/\bar m^2$.  Now we substitute for $a = \Sigma_2 -
\gamma$ in (\ref{eqn:b}) and (\ref{eqn:c}) to find
\beq \label{eqn:d}
\Sigma_1 - 2 N_1 + \Sigma_2 = 2(1+2r) \left( \frac{\beta \Delta}{\bar m}
+ \frac{\gamma}{3} \right)~~,~~~
\Sigma^*_1 - 2 N_1 + \Sigma_2 = 2(1-r) \left( \frac{\beta \Delta}{\bar m}
+ \frac{\gamma}{3} \right)~~~.
\eeq
These two combinations are proportional to one another, so that instead of
being able to solve for $\beta \Delta$ and $\gamma$ we actually have another
mass relation:
\beq \label{eqn:e}
\frac{\Sigma^*_1 - 2 N_1 + \Sigma_2}{\Sigma_1 - 2 N_1 + \Sigma_2} =
\frac{1-r}{1+2r} = 0.14~~~,
\eeq
where we have used $r = 0.675$.  In substituting experimental values in the
left-hand side of this relation, the error on $\Sigma^*_1$ may be reduced by
averaging the direct measurement $\Sigma^*_1 = -4.4 \pm 0.64$ MeV$/c^2$ with
the value $\Sigma^*_1 = \Xi^*_1 + N_1 = - 4.49 \pm 0.6$ MeV$/c^2$ to obtain
$\Sigma^*_1 = -4.46 \pm 0.44$ MeV/$c^2$.  The result for the left-hand side is
$0.04 \pm 0.13$.  The predicted value for $\Sigma^*_1$ is $-4.82 \pm 0.15$
MeV/$c^2$, with the dominant error stemming from $\Sigma_2$.  The corresponding
prediction for $\Xi^*_1 = \Sigma^*_1 - N_1$ is $-3.52 \pm 0.15$ MeV/$c^2$. A
violation of this relation could signify (a) the presence of significant
three-body interactions, (b) a ratio $r$ different from that quoted above, or
(c) the non-negligibility of kinetic two-body terms.  If these are restored,
both the numerator and denominator of the left-hand side of Eq.~(\ref{eqn:e})
involve the combination $2(K_{us} - K_{ds} + K_{dd} - K_{ud})$, which is of
order $\Delta \times$ [SU(3) breaking] and thus is not likely to be
appreciable. 

If the dependence of the kinetic terms on $\Delta$ could be established, one
would have an additional constraint, from which the parameters could be
determined.  As one example, if one totally neglected both the kinetic one- and
two-body mass splittings, one could use $N_1$, $\Sigma_1$, and $\Sigma_2$ to
find $\Delta = -2.57$ MeV$/c^2$, $a = 3.06$ MeV$/c^2$, and $\gamma = -1.35$
MeV$/c^2$.  This value of $\Delta$ is to be compared with that obtained
\cite{models} by Isgur ($-6$ MeV/$c^2$), Capstick ($-4.4$ MeV/$c^2$), Itoh
\ite~($-3.8$ MeV/$c^2$), and Franklin and Lichtenberg ($-2.8$ MeV/$c^2$).
(These last authors point out that Isgur's difference of one-body terms {\em
including kinetic energies} is $-3.0$ MeV/$c^2$, much closer to their (and our)
value of $\Delta$.)  Our Coulomb term $a$ and electromagnetic hyperfine
term $\gamma$ are rather similar to Capstick's.  If the effective potential
between light quarks corresponds to a power-law $r^\nu$ with $\nu > 0$, as is
appropriate for light quarks \cite{QR}, the scaling law mentioned earlier would
imply $K_u > K_d$ for $\Delta < 0$, and hence the inequality $\Delta < -2.57$
MeV/$c^2$, which is satisfied in the models of Ref.~\cite{models}. 
\newpage

\centerline{\bf V.  SYMMETRY VIOLATIONS AND THEIR HIERARCHIES}
\bigskip

\leftline{\bf A.  Non-universality of wave functions}
\bigskip

We assumed universal values of $\langle 1/r_{ij} \rangle$ and
$|\Psi(0)_{ij}|^2$ in deriving the quark-model results of Sec.~II. We found
that hyperfine interactions satisfied our mass relations independently of quark
masses, indicating that we never needed to assume equality of the hyperfine
interaction between two nonstrange quarks and that between a strange and a
nonstrange quark. (The hyperfine interaction between two strange quarks never
entered into our discussion of $\Delta I\ge 1$ mass relations.) However, we did
have to assume that hyperfine interactions between members of a pair were
independent of the environment in which these interactions occurred.  This
assumption was equivalent to the neglect of three-body effects. 

Similarly, we did not have to assume the equality of Coulomb interactions
between non-strange and strange quarks, but had to assume that these
interactions were independent of the environment in which they took place.  To
illustrate this, let us consider the $a$ terms in the Coleman-Glashow relation
(\ref{eqn:CG}).  We shall label the $a$ contributions by subscripts indicating
the interaction quark pair ($n$ for a nonstrange quark and $s$ for a strange
quark) and by a superscript denoting the particle in which the interaction is
taking place.  Then the Coulomb contribution to $N_1 + \Xi_1$ is $(a_{nn}^N - 2
a_{ns}^{\Xi})/3$, while that to $\Sigma_1$ is $(a_{nn}^\Sigma - 2
a_{ns}^\Sigma)/3$.  The only way in which these two terms could differ is if
two-body forces depended on their environment, a circumstance equivalent to the
presence of three-body effects. 

As long as isospin-violating effects are strictly of one-body or two-body
nature, all the relations we have derived so far will hold. What would be a
likely direction for deviations from this circumstance? In the case of the
Coulomb interactions illustrated above, we might expect by considering the
relative size of reduced-mass effects that a nonstrange pair in the $\Sigma$
would be more deeply bound than a nonstrange pair in a nucleon, and a
nonstrange-strange pair in the $\Xi$ more deeply bound than one in a $\Sigma$. 
In that case we would expect 
\beq
a_{nn}^\Sigma - 2 a_{ns}^\Sigma > a_{nn}^N - 2 a_{ns}^{\Xi}~~~,
\eeq
or $\Sigma_1 - (N_1 + \Xi_1) > 0$.  The central value of this relation
is in fact less than zero but with large uncertainty.

A similar ordering of effects holds for the strong hyperfine terms, with
two-body terms contributing with the same relative signs as in the
Coulomb-interaction example.  Thus one expects the same sign of the inequality
from these terms.  On the other hand, the electromagnetic hyperfine
contributions to $N_1$, $\Xi_1$, and $\Sigma_1$ all turn out to be positive,
preventing one from making such an argument.  The two-body kinetic terms
$K_{q_i q_j}$ are of indefinite sign unless one interprets them in a specific
context, e.g., as reduced-mass effects.  Thus if it is ever found that
$\Sigma_1 - (N_1 + \Xi_1) < 0$, a culprit within the quark model may be
three-body effects in kinetic terms or in electromagnetic hyperfine
interactions.  One would have to examine specific models in more detail to see
if such effects really were important. 
\bigskip

\leftline{\bf B.  Comparison with $1/N_c$ hierarchy}
\bigskip

Jenkins and Lebed \cite{JL} have presented a view of isospin-violating mass
splittings, based on a systematic expansion in powers of isospin-breaking and
SU(3)-breaking terms and powers of $1/N_c$.  It is worth reviewing some of the
common points and differences with respect to our approach. 

(1) The $1/N_c$ approach is completely general, whereas we are seeking
interpretations within the quark model. 

(2) Jenkins and Lebed expect the Coleman-Glashow relation to be very good.  A
reduction of errors on $\Xi_1$ by a factor of 2 (to $\pm 0.3$ MeV/$c^2$) should
still lead to a relation which is satisfied to about a standard deviation.  We
are unable to make as quantitative a statement, having not estimated three-body
effects. 

(3) Within the $1/N_c$ approach, certain relations are expected to hold to
better accuracy than others, and the hierarchy does not always agree with that
associated with the number of interacting quarks. 

In the $1/N_c$ approach the $\Delta I = 1$ relations
\beq \label{eqn:twoa}
35[N_1 - \Xi_1 + 2 \sqrt{3} M(\Lambda \Sigma^0)]
- 2(\Delta_1 - 3 \Sigma^*_1 - 4 \Xi^*_1) = 0
\eeq
and
\beq \label{eqn:threea}
7[N_1 - \Xi_1 + 2 \sqrt{3} M(\Lambda \Sigma^0)]
- (\Delta_1 - 3 \Sigma^*_1 - 4 \Xi^*_1) = 0
\eeq
are both expected to hold with the same accuracy, though the first is based on
the suppression of a two-body operator and the second is based on the
suppression of a three-body operator.  In our approach only the second relation
holds.  Other relations based on three-body operators, which consequently hold
in both approaches, and which are expected to be of comparable accuracy to the
first two, are the Coleman-Glashow relation $N_1 - \Sigma_1 + \Xi_1 = 0$ and 
\beq
-7 N_1 - 5 \Sigma_1 + 2 \Xi_1 + 6 \sqrt{3} M(\Lambda \Sigma^0)
+ \Delta_1 + 2 \Sigma^*_1 + \Xi_1^* = 0~~~.
\eeq
The (three-body) relation $\Delta_1 = 10(\Sigma^*_1 - \Xi^*_1)$ is expected to
be better-obeyed by an order of magnitude than the above expressions. Combining
only the three-body relations, we obtain our previous $\Delta I = 1$ results. 
Including Eq.~(\ref{eqn:twoa}), we obtain the additional results 
\beq
N_1 - \Xi_1 + 2 \sqrt{3} M(\Lambda \Sigma^0) = 0~~,~~~
\Delta_1 - 3 \Sigma^*_1 - 4 \Xi^*_1 = 0~~~.
\eeq
When combined with previous results, these imply such relations as
\beq
M(\Lambda \Sigma^0) = (2 \sqrt{3})^{-1}(\Xi_1 - N_1) = - 1.47
\pm 0.17~{\rm MeV}/c^2
\eeq
(as quoted in Ref.~\cite{JL}), and, eliminating $M(\Lambda \Sigma^0)$ from the
above relation and Eq.~(\ref{eqn:arb}), $\Sigma_1 - \Sigma^*_1 = \Xi_1 - N_1$,
or, using other relations expected to hold to the same order, 
\beq
\Sigma_1^*/2~(= -2.2 \pm 0.32~{\rm MeV}/c^2)
= \Xi_1^*~(= -3.2 \pm 0.6~{\rm MeV}/c^2)
= N_1~(= -1.293 ~{\rm MeV}/c^2)~~~.
\eeq
The result $\Sigma^*_1/2 = \Xi^*_1$, as noted in Ref.~\cite{JL}, is obeyed to
$0.04 \pm 0.03\%$.  The definition of accuracy adopted there, which we shall
use, involves writing mass relations as (LHS) = (RHS), with all terms positive;
accuracy is then defined as (LHS -- RHS)/[(LHS + RHS)/2]. The relation
$\Sigma^*_1/2 = N_1$ is obeyed to $-0.06 \pm 0.02\%$, while $\Xi^*_1 = N_1$
holds to $-0.077 \pm 0.024\%$.  All these relations are expected to hold to
about $\pm 0.03\%$. Thus the results involving $N_1$ correspond to slightly
worse accuracy than expected, but not at a significant level.  Improvement of
experimental accuracy on $\Sigma^*_1$ and $\Xi^*_1$ would be very helpful in
testing the predicted hierarchy. 

We mentioned previously a relation obtained by combining $\Delta_1 = 10 N_1$
with $\Delta_3 = 0$, namely $M(\Delta^+) - M(\Delta^0) = N_1$. This relation is
expected in Ref.~\cite{JL} to be good to $\pm 0.03\%$, while it is observed to
$0.11 \pm 0.06\%$.  Improved information on the $\Delta^+$ mass would be needed
to test this result significantly.  Another such relation is $M(\Delta^{++} -
M(\Delta^-) = 3 N_1$, totally untested at present.

In a study of $\Delta I = 2$ relations, the $1/N_c$ hierarchy appears to have
more success than our neglect of three-body effects. The two-body relation 
\beq \label{eqn:twob}
35 \Sigma_2 - 2 (3 \Delta_2 + \Sigma_2^*) = 0
\eeq
and the three-body relation
\beq \label{eqn:threeb}
7 \Sigma_2 - (3 \Delta_2 + \Sigma_2^*) = 0
\eeq
are expected to hold to the same order, whereas we obtain only the second in
the quark model.  The relations are satisfied to $0.10 \pm 0.03\%$ and $0.11
\pm 0.05\%$, respectively.  [We have used $\Delta_3 = 0$ to eliminate
$M(\Delta^-)$.  The authors of Ref.~\cite{JL} obtain slightly different results
as a result of different $M(\Delta)$ inputs.] Both approaches obtain the
three-body relation $\Delta_2 = 2 \Sigma_2^*$, which is expected in the $1/N_c$
analysis to be an order of magnitude more accurate than (\ref{eqn:twob}) or
(\ref{eqn:threeb}). 

Combining (\ref{eqn:twob}) and (\ref{eqn:threeb}), we find that to the same
accuracy, 
\beq
\Sigma_2~(= 1.71 \pm 0.18~{\rm MeV}/c^2) = 0~~,~~~
\Delta_2 = - \Sigma_2^*/3~~~.
\eeq
The result $\Sigma_2 = 0$ is good to 0.07\%.  This may serve as a benchmark for
the accuracy to which (\ref{eqn:twob}) and (\ref{eqn:threeb}) may be expected
to hold.  The result $\Delta_2 = - \Sigma_2^*/3$, when combined with $\Delta_2
= 2 \Sigma_2^*$ (expected to be more accurate), implies $\Delta_2 = 0$ and
$\Sigma^*_2 = 0$. $\Delta_2 = 0$, when combined with $\Delta_3 = 0$, implies 
\beq
M(\Delta^{++}) - 2 M(\Delta^+) + M(\Delta^0)~(= -5.0 \pm 2.8
~{\rm MeV}/c^2) = 0~~~,
\eeq
corresponding to an error of $-0.2 \pm 0.1\%$. This is slightly better than the
corresponding relation (\ref{eqn:dsb}) in our approach (when we use $\Sigma_2$
on the right-hand side.) The predicted relation $\Sigma_2^*~(= 2.6 \pm 2.1~{\rm
MeV}/c^2) = 0$ is obeyed to $0.09 \pm 0.08\%$. The poorly obeyed relation
(\ref{eqn:dsa}) is replaced by 
\beq
M(\Delta^{++}) - M(\Delta^0)~(= -2.7 \pm 0.3~{\rm MeV}/c^2) = 2 N_1~ 
(= -2.6~{\rm MeV}/c^2)~~~,
\eeq
obeyed to $-0.005 \pm 0.014\%$; the expected accuracy, however, is only as good
as that for $\Sigma_2 = 0$, i.e., $\pm 0.07\%$. 

The $1/N_c$ hierarchy suggests that the relation $\Delta_2 = 2 \Sigma_2^*$
should be a factor ${\cal O}(\epsilon/N_c)$ more accurate than (\ref{eqn:twob})
or (\ref{eqn:threeb}), where $\epsilon \sim 1/4$ describes SU(3) breaking. 
Thus, we could expect it to hold to $\pm 0.01\%$ (or at worst $\pm 0.02\%$
using the numerical estimates of Ref.~\cite{JL}).  Using the relation $\Delta_3
= 0$ whose errors are negligible by comparison, we find the ensuing relation 
\beq
M(\Delta^{++}) -2 M(\Delta^+) + M(\Delta^0)~(= -5.0 \pm 2.8~{\rm MeV}
/c^2) = \Sigma^*_2~(= 2.6 \pm 2.1~{\rm MeV}/c^2)
\eeq
to be satisfied to $-0.15 \pm 0.07\%$.  Reduction of the errors on
$M(\Delta^+)$ and $M(\Sigma^{*0})$ is necessary to perform an incisive test of
this result.  If we use the relation $M(\Delta^+) = N_1 + M(\Delta^0)$,
expected to be good to $\pm 0.03\%$ as mentioned above, we find that
$M(\Delta^{++}) - M(\Delta^0) = \Sigma^*_2 + 2 N_1$ [Eq.~(\ref{eqn:dsa}) with
$\Sigma^*_2$] should be satisfied to $\pm 0.03\%$, whereas it holds to $-0.05
\pm 0.04\%$.  Here the error is dominated by that of $M(\Sigma^{*0})$.
As noted earlier, some of these conclusions will be changed if the $\Delta^+$
mass quoted in Ref.~\cite{PDG} is found to be in error \cite{Deltas}.
\bigskip

\centerline{\bf VI.  CHARMED BARYONS}
\bigskip

Although baryons containing light quarks are the present topic,
recent works \cite{JF,Varga} have noted a serious discrepancy between
a $\Delta I = 2$ relation between charmed and non-charmed baryons.  The
assumptions of Sec.~II lead to the relation \cite{JFearly}
\beq
\Sigma_{c2}~(= -2.0 \pm 1.3~{\rm MeV}/c^2) = \Sigma_2~(= 1.71 \pm 0.18~
{\rm MeV}/c^2)~~~,
\eeq
where $\Sigma_{c2} \equiv M(\Sigma_c^{++}) - 2 M(\Sigma_c^+) + M(\Sigma_c^0)$.
Experimental values are taken from Ref.~\cite{PDG}. As noted by \cite{JF}, the
negative sign of $\Sigma_{c2}$ is very difficult to understand in the quark
model. 

One $\Delta I =1$ relation \cite{JFearly} follows from our assumptions:
\beq
\Sigma_{c1} - 2 \Xi'_{c1} = \Sigma^*_1 - 2 \Xi^*_1~~~,
\eeq
where $\Sigma_{c1} \equiv M(\Sigma_c^{++}) - M(\Sigma_c^0) = 0.8 \pm 0.4~{\rm
MeV}/c^2$ \cite{PDG} and $\Xi'_{c1} \equiv M({\Xi'_c}^+) - M({\Xi'_c}^0) = -1.7
\pm 4.6~{\rm MeV}/c^2$ \cite{CLEOXi}.  (Here $\Xi'_c$ denotes the state in
which the light quarks are in a flavor- and spin-symmetric state.) The large
error on the last quantity prevents any test at present. 
\bigskip

\centerline{\bf VII.  SUMMARY}
\bigskip

The prospect of improved values of masses for baryons, such as $\Xi^0$ in the
KTeV Experiment at Fermilab, charged hyperons and perhaps $\Sigma^*$'s and
$\Xi^*$'s in other Fermilab experiments, and $\Delta$'s in high-intensity
photoproduction studies, has led us to re-examine predictions for isospin
splittings within the assumption of one- or two-body effects within the quark
model.  We have shown that the Coleman-Glashow relation (\ref{eqn:CG}) is
expected to be satisfied independently of quark masses within this assumption.
A deviation from it would have to be ascribed to three-body effects. 

Tests of other relations will require improved knowledge of decuplet isospin
splittings. (For notation see Sec.~II.) These include the $\Delta I = 1$
relations $N_1 = \Delta_1/10 = \Sigma^*_1 - \Xi^*_1$ and $2 \sqrt{3} M(\Lambda
\Sigma_0) = \Sigma_1 - \Sigma^*_1$, the $\Delta I = 2$ relations $\Sigma_2 =
\Sigma^*_2 = \Delta_2/2$, and the relation $\Delta_3 = 0$. 

We have discussed the degree to which one can isolate individual contributions
to mass splittings, given quark masses obtained in fits to baryon octet and
decuplet spectra \cite{GR}.  A model-independent determination of these
parameters is not possible as a result of the presence of one- and two-body
kinetic energy terms.  Under a restricted set of assumptions
one obtains the predictions $\Sigma^*_1 = - 4.82 \pm 0.15$ MeV/$c^2$
and $\Xi^*_1 = - 3.52 \pm 0.15$ MeV/$c^2$.

With dynamical assumptions, it would be possible to estimate kinetic terms. One
could thereby solve for quantities such as the intrinsic constituent-quark mass
difference $\Delta = m_u - m_d$, the Coulomb self-energy, and the
electromagnetic hyperfine interaction.  Knowledge of $\Delta$ would be useful
in evaluating heavy meson decay constants using spin-dependent hyperfine
interactions in the $D$ and $D^*$ systems \cite{FM}.  Knowledge of the
electromagnetic hyperfine interaction term $\gamma$, providing an estimate of
$|\Psi_{ij}(0)|^2$ in Eq.~(\ref{eqn:HFe}), would be useful in calculations of
nonleptonic weak decays of hyperons \cite{LeY} or of proton decay \cite{PDK}. 

We have compared our approach with that of a systematic $1/N_c$ expansion
\cite{JL}, where $N_c$ is the number of colors in QCD. The presence of two- and
three-body operators of similar order in the $1/N_c$ expansion leads to a
hierarchy of mass relations somewhat different from ours.  That approach
suggests that the Coleman-Glashow relation should be good to about $\pm 0.3$
MeV/$c^2$ or better.  The $1/N_c$ expansion also obtains some $\Delta I = 1$
relations which are expected to be as good as the Coleman-Glashow relation,
such as $\Sigma_1^*/2 = \Xi^*_1 = N_1$.  It will be interesting to compare
violations of these and several $\Delta I = 2$ relations with that of the
Coleman-Glashow relation once improved data on decuplet isospin splittings are
available. 
\bigskip

\centerline{\bf ACKNOWLEDGMENTS}
\bigskip

I wish to thank A. Dighe, J. Franklin, E. Jenkins, A. Manohar, N. Solomey, E.
Swallow, and R. Winston for discussions, and E. Jenkins for a careful reading
of the manuscript.  This work was performed in part at the Aspen Center for
Physics and supported in part by the United States Department of Energy under
Grant No. DE FG02 90ER40560. 
\newpage

\def \ajp#1#2#3{Am. J. Phys. {\bf#1}, #2 (#3)}
\def \apny#1#2#3{Ann. Phys. (N.Y.) {\bf#1}, #2 (#3)}
\def \app#1#2#3{Acta Phys. Polonica {\bf#1}, #2 (#3)}
\def \arnps#1#2#3{Ann. Rev. Nucl. Part. Sci. {\bf#1}, #2 (#3)}
\def \cmts#1#2#3{Comments on Nucl. Part. Phys. {\bf#1}, #2 (#3)}
\def \cn{Collaboration}
\def \cp89{{\it CP Violation,} edited by C. Jarlskog (World Scientific,
Singapore, 1989)}
\def \dpfa{{\it The Albuquerque Meeting: DPF 94} (Division of Particles and
Fields Meeting, American Physical Society, Albuquerque, NM, Aug.~2--6, 1994),
ed. by S. Seidel (World Scientific, River Edge, NJ, 1995)}
\def \dpff{{\it The Fermilab Meeting: DPF 92} (Division of Particles and Fields
Meeting, American Physical Society, Batavia, IL., Nov.~11--14, 1992), ed. by
C. H. Albright \ite~(World Scientific, Singapore, 1993)}
\def \efi{Enrico Fermi Institute Report No. EFI}
\def \epl#1#2#3{Europhys.~Lett.~{\bf #1}, #2 (#3)}
\def \f79{{\it Proceedings of the 1979 International Symposium on Lepton and
Photon Interactions at High Energies,} Fermilab, August 23-29, 1979, ed. by
T. B. W. Kirk and H. D. I. Abarbanel (Fermi National Accelerator Laboratory,
Batavia, IL, 1979}
\def \hb87{{\it Proceeding of the 1987 International Symposium on Lepton and
Photon Interactions at High Energies,} Hamburg, 1987, ed. by W. Bartel
and R. R\"uckl (Nucl. Phys. B, Proc. Suppl., vol. 3) (North-Holland,
Amsterdam, 1988)}
\def \ib{{\it ibid.}~}
\def \ibj#1#2#3{~{\bf#1}, #2 (#3)}
\def \ichep72{{\it Proceedings of the XVI International Conference on High
Energy Physics}, Chicago and Batavia, Illinois, Sept. 6 -- 13, 1972,
edited by J. D. Jackson, A. Roberts, and R. Donaldson (Fermilab, Batavia,
IL, 1972)}
\def \ijmpa#1#2#3{Int. J. Mod. Phys. A {\bf#1}, #2 (#3)}
\def \jpb#1#2#3{J.~Phys.~B~{\bf#1}, #2 (#3)}
\def \lkl87{{\it Selected Topics in Electroweak Interactions} (Proceedings of
the Second Lake Louise Institute on New Frontiers in Particle Physics, 15 --
21 February, 1987), edited by J. M. Cameron \ite~(World Scientific, Singapore,
1987)}
\def \ky85{{\it Proceedings of the International Symposium on Lepton and
Photon Interactions at High Energy,} Kyoto, Aug.~19-24, 1985, edited by M.
Konuma and K. Takahashi (Kyoto Univ., Kyoto, 1985)}
\def \mpla#1#2#3{Mod. Phys. Lett. A {\bf#1}, #2 (#3)}
\def \nc#1#2#3{Nuovo Cim. {\bf#1}, #2 (#3)}
\def \np#1#2#3{Nucl. Phys. {\bf#1}, #2 (#3)}
\def \pisma#1#2#3#4{Pis'ma Zh. Eksp. Teor. Fiz. {\bf#1}, #2 (#3) [JETP Lett.
{\bf#1}, #4 (#3)]}
\def \pl#1#2#3{Phys. Lett. {\bf#1}, #2 (#3)}
\def \pla#1#2#3{Phys. Lett. A {\bf#1}, #2 (#3)}
\def \plb#1#2#3{Phys. Lett. B {\bf#1}, #2 (#3)}
\def \pr#1#2#3{Phys. Rev. {\bf#1}, #2 (#3)}
\def \prc#1#2#3{Phys. Rev. C {\bf#1}, #2 (#3)}
\def \prd#1#2#3{Phys. Rev. D {\bf#1}, #2 (#3)}
\def \prl#1#2#3{Phys. Rev. Lett. {\bf#1}, #2 (#3)}
\def \prp#1#2#3{Phys. Rep. {\bf#1}, #2 (#3)}
\def \ptp#1#2#3{Prog. Theor. Phys. {\bf#1}, #2 (#3)}
\def \ptwaw{Plenary talk, XXVIII International Conference on High Energy
Physics, Warsaw, July 25--31, 1996}
\def \rmp#1#2#3{Rev. Mod. Phys. {\bf#1}, #2 (#3)}
\def \rp#1{~~~~~\ldots\ldots{\rm rp~}{#1}~~~~~}
\def \si90{25th International Conference on High Energy Physics, Singapore,
Aug. 2-8, 1990}
\def \slc87{{\it Proceedings of the Salt Lake City Meeting} (Division of
Particles and Fields, American Physical Society, Salt Lake City, Utah, 1987),
ed. by C. DeTar and J. S. Ball (World Scientific, Singapore, 1987)}
\def \slac89{{\it Proceedings of the XIVth International Symposium on
Lepton and Photon Interactions,} Stanford, California, 1989, edited by M.
Riordan (World Scientific, Singapore, 1990)}
\def \smass82{{\it Proceedings of the 1982 DPF Summer Study on Elementary
Particle Physics and Future Facilities}, Snowmass, Colorado, edited by R.
Donaldson, R. Gustafson, and F. Paige (World Scientific, Singapore, 1982)}
\def \smass90{{\it Research Directions for the Decade} (Proceedings of the
1990 Summer Study on High Energy Physics, June 25--July 13, Snowmass, Colorado),
edited by E. L. Berger (World Scientific, Singapore, 1992)}
\def \tasi90{{\it Testing the Standard Model} (Proceedings of the 1990
Theoretical Advanced Study Institute in Elementary Particle Physics, Boulder,
Colorado, 3--27 June, 1990), edited by M. Cveti\v{c} and P. Langacker
(World Scientific, Singapore, 1991)}
\def \waw{XXVIII International Conference on High Energy
Physics, Warsaw, July 25--31, 1996}
\def \yaf#1#2#3#4{Yad. Fiz. {\bf#1}, #2 (#3) [Sov. J. Nucl. Phys. {\bf #1},
#4 (#3)]}
\def \zhetf#1#2#3#4#5#6{Zh. Eksp. Teor. Fiz. {\bf #1}, #2 (#3) [Sov. Phys. -
JETP {\bf #4}, #5 (#6)]}
\def \zpc#1#2#3{Zeit. Phys. C {\bf#1}, #2 (#3)}
\def \zpd#1#2#3{Zeit. Phys. D {\bf#1}, #2 (#3)}


\begin{thebibliography}{99}
\bibitem{Isgur} The Coulomb term for the mass difference $M_{\Sigma^-} -
M_{\Sigma^0}$ in Table I of the study by N.~Isgur \cite{PostQCD} should be
$2 \epsilon/3$, not $\epsilon/3$.
\end{thebibliography}
\end{document}